\documentclass[amsmath,amssymb,prl,hyperlink,twocolumn]{revtex4}

	\usepackage{graphicx}
	\usepackage{soul}
	\usepackage[colorlinks=true,citecolor=blue,linkcolor=magenta]{hyperref}
	\usepackage[usenames]{color}
	\usepackage{amsfonts}
	\usepackage{color}
	\usepackage{booktabs}
	\usepackage{multirow}
	\usepackage{float}
    \usepackage{times}
    \usepackage[english]{babel}
     \usepackage{float}

\begin{document}

\title{99.9\%-fidelity in measuring a superconducting qubit}
\author{Can Wang$^{1,2,3,*}$, 
Feng-Ming Liu$^{1,2,3,*}$, 
He Chen$^{1,2,3}$, 
Yi-Fei Du$^{1,2,3}$,
Chong Ying$^{1,2,3}$,
Jian-Wen Wang$^{1,2,3}$,
Yong-Heng Huo$^{1,2,3}$,
Cheng-Zhi Peng$^{1,2,3}$,
Xiaobo Zhu$^{1,2,3}$,
Ming-Cheng Chen$^{1,2,3,\dagger}$,
 Chao-Yang Lu$^{1,2,3,\dagger}$, and Jian-Wei Pan$^{1,2,3,\dagger}$ \vspace{0.2cm}}

\affiliation{$^1$Hefei National Research Center for Physical Sciences at the Microscale and Department of Modern Physics, University of Science and Technology of China, Hefei, Anhui 230026, China}
\affiliation{$^2$Shanghai Research Center for Quantum Science and CAS Center for Excellence in Quantum Information and Quantum Physics, University of Science and Technology of China, Shanghai 201315, China}
\affiliation{$^3$Hefei National Laboratory, University of Science and Technology of China, Hefei 230088, China}
\date{\today}

\begin{abstract}
	Despite the significant progress in superconducting quantum computation over the past years,
	quantum state measurement still lags nearly an order of magnitude behind quantum gate operations in speed and fidelity.
	 The main challenge is that the strong coupling and readout signal used to probe the quantum state may also introduce additional
	channels which may cause qubit state transitions. 
	Here, we design a novel architecture to implement the long-sought longitudinal interaction scheme between qubits and resonators. 
	This architecture not only provides genuine longitudinal interaction by eliminating residual transversal couplings, 
	but also introduces proper nonlinearity to the resonator that can further minimize decay error and measurement-induced excitation error. 
	Our experimental results demonstrate a measurement fidelity of 99.8\% in 202 ns without the need for any first-stage amplification. 
	After subtracting the residual preparation errors, the pure measurement fidelity is above 99.9\%. 
	Our scheme is compatible with the multiplexing readout scheme and can be used for quantum error correction.
\end{abstract}

\maketitle
Dispersive measurement\cite{1} has become a widely used non-demolition readout protocol for superconducting qubits in quantum computation\cite{2}, 
simulation\cite{3,4}, and error correction\cite{5,6,7}. 
Linear readout resonators connect the superconducting qubits to the environment, 
serving as mediators to acquire the quantum state without directly manipulating the qubit. 
A key challenge in improving the fidelity of dispersive measurement is to obtain high signal-to-noise ratio in a time much shorter than the qubit lifetime. 
To address this, it is essential to enhance the dispersive coupling strength without introducing additional leakage channels, 
such as the Purcell-enhanced relaxation\cite{1} and measurement-induced excitation\cite{8}.
A variety of designs in both hardware architecture and control method have been developed to improve the measurement fidelity. 
First, microwave amplifiers at multiple temperature stages are applied to amplify the readout signal\cite{9,10,11,12}. The multilevel readout protocol pre-excited the first excited state to higher energy levels\cite{13,14} and the Josephson bifurcation amplifier\cite{15} 
used nonlinear resonators to decouple qubit from readout resonators during integration which reduce the qubit decay error from software 
and architecture level respectively. 
Moreover, Purcell filters are developed to alleviate the tradeoff between Purcell-enhanced qubit relaxation and the exchanging rate between the 
readout resonator and the environment\cite{16,17,18}. 
These techniques partially but not completely addressed the limitations of dispersive measurement. 
In general, quantum state measurement is still the weakest part in quantum computation.
In this work, we propose and demonstrate a novel design to be implemented the long-sought longitudinal interaction scheme. 
It can simultaneously reduce the Purcell-enhanced decay rate, the qubit decay error and measurement-induced qubit transitions. 
In our design, the Transmon qubit\cite{19,20} and the transmission-line-based (TL) readout resonator are coupled via a Josephson junction. 

The introduction of the coupling Josephson junction has two main advantages. First, 
it can generate genuine longitudinal interaction between the qubit and the resonator, 
in contrast to the capacitive coupling designs which are transversal interactions. 
Compared with the previous longitudinal coupling proposals which are not yet realized in experiment\cite{21,22}, 
our design is much easier to implement as it keeps the basic modular components and architecture of traditional superconducting quantum processors. 
Second, the Josephson junction also adds weak nonlinearity to the TL resonator. 
In addition to utilizing the bifurcation nature of nonlinear resonators to suppress the qubit decay error,
 we further identified that the asymmetric response of nonlinear resonators in the frequency domain could be used to enhance the readout power. 

\begin{figure*}[tb]
	\centering
	\includegraphics[width=0.9\textwidth]{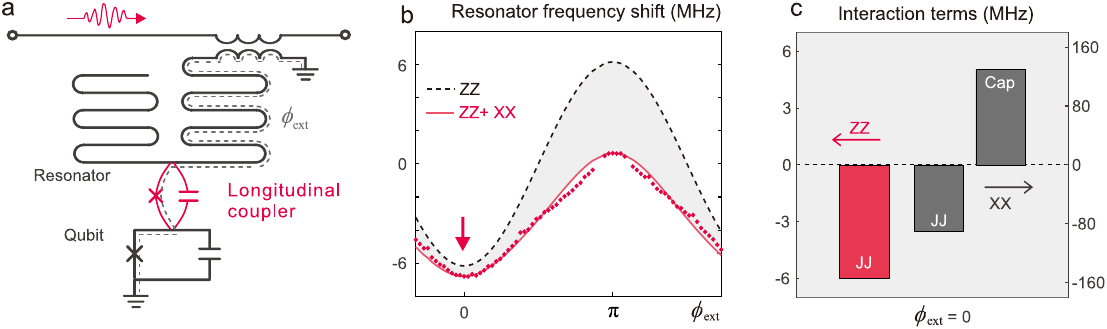}
	\caption{\textbf{Realization of genuine longitudinal coupling.} 
	\textbf{(a)} Schematic of the sample architecture: a quarter-wavelength resonator is coupled to the qubit through a Josephson junction for readout. 
	\textbf{(b)} The resonator’s dispersive shift as a function of external flux. The dashed line and the shaded area represent the contribution of ZZ and XX 
	interaction, respectively. 
	\textbf{(c)} Illustration of interaction terms at $\phi_{\text{ext}} = 0$. The red bar represents the ZZ interaction strength, 
	and the gray bars represent the XX interaction strength from CJJ and mutual capacitance, which have opposite sign. }
	\label{fig1}
\end{figure*}

 We begin by introducing the design architecture and the model Hamiltonian. 
 The structure of our sample is shown in Fig.\ref{fig1}a. 
 A Transmon qubit with anharmonicity $\alpha = -290~\text{MHz}$ is coupled to a quarter-wavelength TL resonator through a Josephson junction, 
 which acts as a longitudinal coupler. The connecting point of the resonator and junction can be varied according to the demanding of parameters.
 In this work, it is in the middle of the resonator. The Josephson energy of the coupling Josephson junction (CJJ) is  $E_{J,C}/h=2.0 ~\text{GHz}$. 
 It is small compared with that of the junction in the qubit, 
 so that we can quantize the resonator and the qubit separately and get their interaction Hamiltonian (Supplemental Materials). 

 After rotating wave approximation, the resulting interaction Hamiltonian is:
\begin{equation}
	\begin{aligned}
		\hat{H}_I=&E_{J C} \cos \phi_{e x t}\left[\frac{1}{2} \Delta_q \hat{\sigma}_z+\Delta_r \hat{a}^{\dagger} \hat{a}+\frac{1}{2} \eta\left(\hat{a}^{\dagger} \hat{a}\right)^2\right.
		\\&+g_{ZZ} \hat{\sigma}_z \hat{a}^{\dagger} \hat{a}+g_{X X}\left(\hat{\sigma}_{+} \hat{a}+\hat{\sigma}_{-} \hat{a}^{\dagger}\right)],
	\end{aligned}
	\label{eq1}
\end{equation}
where $\hat{a}(\hat{a}^\dagger)$ and $\hat \sigma_{-}(\hat \sigma_{+})$ are the annihilation(creation) operators of the resonator and the qubit and
 $\phi_{\text{ext}}$ is the external flux threading the loop labeled by the gray dashed line in Fig.\ref{fig1}a, in unit of the reduced flux quantum $\Phi_0/2\pi=\hbar/2e$. 
 The first term in the square brackets represents the frequency shift of the qubit. 
 The second and the third term represent the resonator’s frequency shift and nonlinearity.
 In our device, when $\phi_{\text{ext}}$ varies from $0$ to $\pi$, 
 the qubit can be tuned from $5.2 ~\text{GHz}$ to $4.3 ~\text{GHz}$, 
 and the resonator can be tuned in a range of $50 ~\text{MHz}$ around $6.66 ~\text{GHz}$. 
 The fourth term represents the longitudinal (ZZ) coupling between the qubit and resonator 
 while the fifth term represents the transversal (XX) coupling. 
 All of them are modulated sinusoidally by the external flux $\phi_{\text{ext}}$. 

 A remarkable feature in our design is that the unexpected transversal (XX) interaction 
 in the Hamiltonian can be eliminated through simple interference to implement genuine 
 longitudinal (ZZ) coupling. In fact, in addition to the Josephson junction coupling, 
 there is also some capacitive coupling between the qubit and resonator. 
 Their mutual capacitance arises from two aspects. The first is the parasitic capacitance in the CJJ, 
 which we estimate is $4  ~\text{fF}$. The second comes from the planar electrode of the qubit. 
 According to our simulation, it is $2.9  ~\text{fF}$. These mutual capacitances can give rise 
 to transversal interaction strength with opposite sign to that originated from the 
 Josephson junction coupling. They interfere destructively at $\phi_{\text{ext}}=0$, and constructively at $\phi_{\text{ext}}=\pi$.

In Fig.\ref{fig1}b, 
we show the interference nature of transversal (XX) interaction in our sample. 
The total coupling strength is characterized through resonator frequency shift 
$f_{\text{C0}}-f_{\text{C1}}$  at different external flux, where $f_{\text{C0}}(f_{\text{C1}})$
is the resonator frequency when the qubit state is  $\left|0\right>(\left|1\right>)$. 
The red dots are measured frequency shifts and the black dashed line is the fitted longitudinal coupling component. 
The longitudinal coupling component is a sinusoidal function which varies from negative to positive when 
the external flux shifts from $0$ to $\pi$.
In contrast, the contribution from transversal coupling is always negative and pulls 
the total shift of the resonator downwards as indicated by the shaded area (see Fig.\ref{fig1}b).

When $\phi_{\text{ext}}=0$, 
the transversal (XX) coupling strength is the smallest (Fig.\ref{fig1}c). 
At this point, 90\% of the dispersive shift is contributed by the genuine longitudinal interaction. 
This alleviates many side-effects of transversal interaction, 
including the dressed dephasing effect in strong readout signal\cite{23}
and Purcell effect which limits the qubit’s relaxation time. 
This allows us to eliminate the need for on-chip Purcell filters in future large-scale integrations. 
Also, this type of couplings gives us the potential to enhance the resonator’s frequency shift beyond the dispersive limit.

After eliminating the transversal coupling, 
we proceed to exploit the nonlinear steady states of the resonator for qubit measurement. 
As described in Eq.\ref{eq1},
the CJJ also adds some nonlinearity to the resonator\cite{15}. 
By driving the nonlinear resonator, a steady state of resonator that corresponds to initial qubit state $\left|1\right>$
can be held, even if the qubit state subsequently decays to state $\left|0\right>$.
This ensures that the state of the resonator is only sensitive to the initial state of the qubit, 
but not the qubit state after the resonator has been established. 
This decoupling of the resonator from the qubit removes the qubit decay error during quantum measurement. 

\begin{figure*}[tb]
	\centering
	\includegraphics[width=0.9\textwidth]{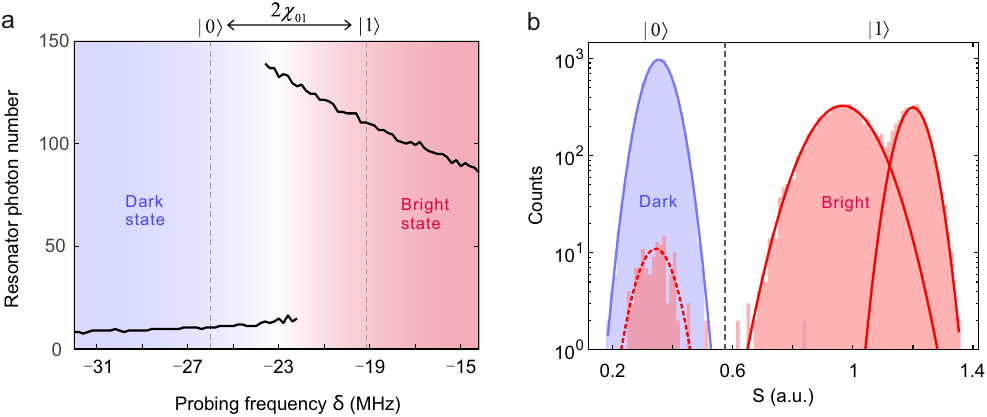}
	\caption{\textbf{Characteristics of the nonlinear resonator.}
	\textbf{(a)} The photon number’s dependence on readout frequency. The continuously driven resonator varies from dark state to bright state as the probing frequency increases and is strongly bifurcated in the middle of the graph. The dashed lines indicate the region where we made our measurements.
	\textbf{(b)} Presentation of the multiple steady states. The readout signal distribution corresponding to qubit state  $\left|0\right>$ and $\left|1\right>$ are shown by blue and red, respectively. }
	\label{fig2}
\end{figure*}

In our measurement, we carefully chose the driving frequency to leverage the multiple steady states
for ultrahigh fidelity measurement. 
In Fig.\ref{fig2}a,
we show the photon number of the resonator depending on the driving frequency\cite{24},
when the qubit is prepared in state $\left|0\right>$. 
There are clearly two distinct branches, which we refer as “bright state” and “dark state”, 
based on their corresponding photon number. 
In the middle of the graph, 
we can observe an overlap of the two branches, 
which means that both of the steady states are excited significantly. 
Hence, quantum measurement operated in this range will yield substantial 
out-of-equilibrium error due to the extra steady states. 
Therefore, in order to get a high measurement fidelity, 
the resonator’s frequency shift should be large enough so that we are away from this region no matter the qubit 
is in state $\left|0\right>$ or  $\left|1\right>$. 

The parameters of our following measurements are indicated by the dashed lines of Fig.\ref{fig2}a. 
Here, when the qubit is in $\left|0\right>$, 
the photons in the resonator are very few, however, when the qubit is in $\left|1\right>$,
because of the change in relative driving frequency,
there are more than ten folds of photons in the resonator. 
In Fig.\ref{fig2}b, we present a detailed view of the measurement signals.

The asymmetry of the photon population holds the $\left|0\right>$ component of the qubit state 
even in the presence of high driving power. 
Although the $\left|1\right>$ component may be altered,
it can only be excited to higher energy levels of the qubit\cite{8},
without affecting our measurement. Combined with a reset protocol\cite{25},
the measurement scheme can be completely non-demolition and is compatible with quantum error corrections. 
This is also confirmed by an active initialization protocol embedded in our experiment\cite{26}. 
In this way, our setup has overcome two main limitations of the superconducting quantum measurement:
decay error and measurement induced excitation error, which paves the way for our ultrahigh fidelity 
quantum measurement. 

\begin{figure*}[tb]
	\centering
	\includegraphics[width=0.9\textwidth]{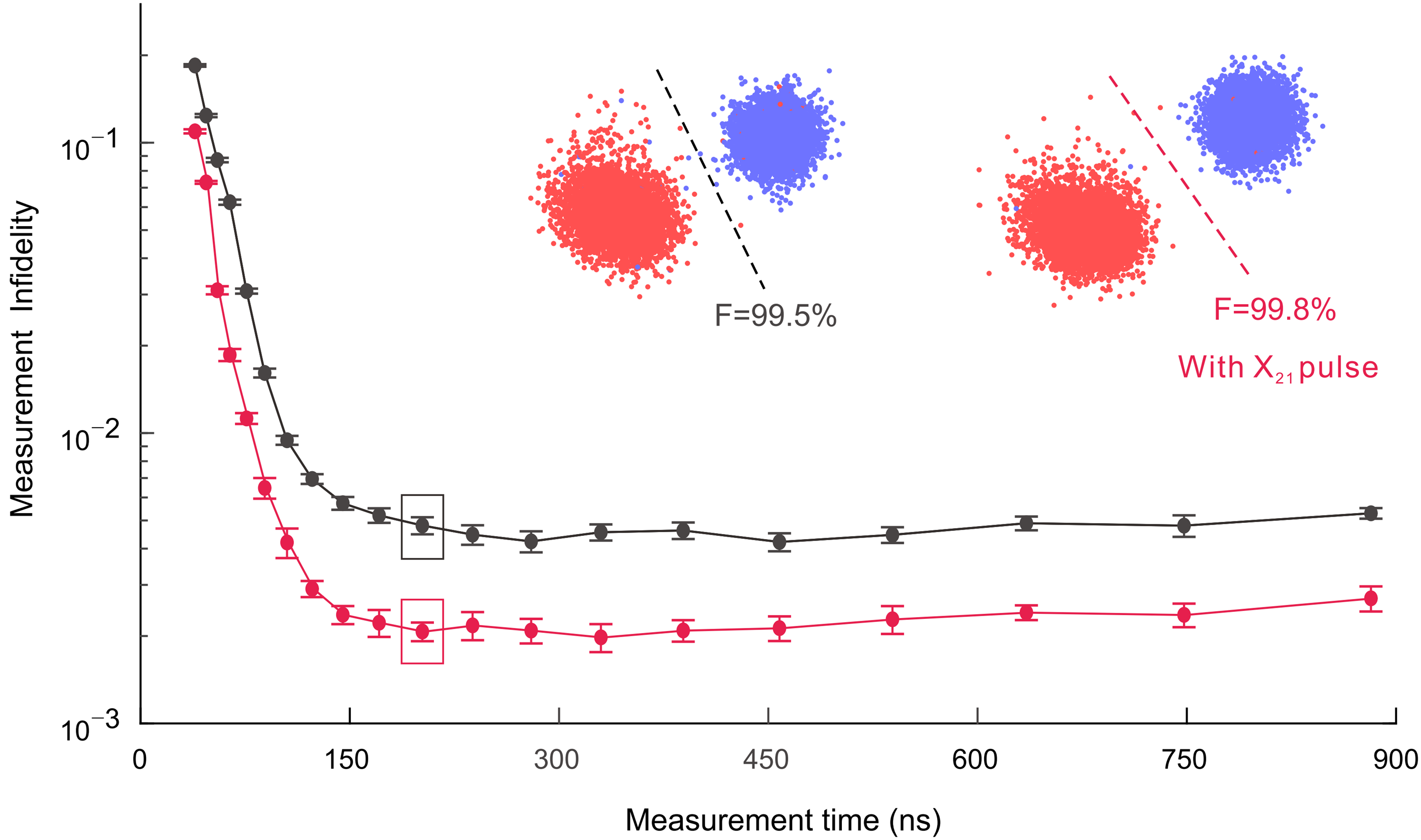}
	\caption{\textbf{Demonstration of the ultrahigh fidelity quantum measurement.}
	The red and black line show the experimental measurement fidelity with and without pre-excitation 
	of the qubit from $\left|1\right>$ to $\left|2\right>$.
	The insets are the IQ distribution of two typical parameters as indicated by the squares. 
	The measurement time here is $202~\text{ns}$.
	}
 	\label{fig3}
\end{figure*}

Finally, we benchmarked the measurement fidelity at zero external flux. 
At this flux point, the qubit’s relaxation and dephasing time 
are $T_1=15~\mu \text{s}$ and $T_2^*=5.6~\mu \text{s}$.
We varied the measurement time from $39~\text{ns}$ to $882~\text{ns}$.
For each point, we optimized the driving frequency and amplitude to minimize the readout error,
and the resulting infidelity is plotted as red dots in Fig.\ref{fig3}. 
Here, the infidelity is defined as $(P(0|1)+P(1|0))/2$, where $P(0|1)$ denotes 
the probability that the qubit is prepared in $\left|1\right>$ but measured to be $\left|0\right>$, and vice versa.
When the measurement time reaches $202~\text{ns}$,
the readout fidelity is 99.5\%. 
To further improve our readout fidelity, 
we apply an X12 gate at the beginning of the measurement to pre-excite the $\left|1\right>$ component to $\left|2\right>$.
This multilevel readout protocol has been previously exploited to alleviate decay error\cite{9,10}. 
In our experiment, it is mainly employed to reduce the out-of-equilibrium error, 
resulting from the larger frequency shift of the readout resonator induced by the higher energy levels 
of the qubit. Compared with the simple discrimination method without pre-excitation, 
our method reduces the average infidelity by approximately 50\%,
as shown by the redline in Fig.\ref{fig3}. 
All of the data are averaged from 10 rounds of experiments, 
each containing 30000 individual measurements. 
The error bars in the graph represent the standard deviations of the results from these 10 rounds. 

As the measurement time increases, 
the separation error decreases and the final infidelity converges to 0.2\%. 
This has surpassed the most state-of-the-art readout result as reported in ref.\cite{27}. 
In fact, it has also accounted in some considerable preparation errors. 
Although we have used a heralding measurement to subtract the qubit’s thermal excitation, 
we estimate that about 0.1\% population of excited state can be accumulated in the $1~\mu\text{s}$ 
vacancy after the heralding measurement (see SM). 
The non-ideal X gate from $\left|0\right>$  to $\left|1\right>$ is also a source of preparation error. 
Therefore, we estimate the pure measurement fidelity to be larger than 99.9\%.

The high readout fidelity comes from two key factors. 
First, the system generates genuine longitudinal coupling, 
which provides large frequency shift of the resonator while eliminating the unexpected transversal (XX) interaction. 
The second comes from the steady-state feature of nonlinear resonators, 
including the suppression of decay error and measurement induced excitation error. 
Notably, we didn’t employ any first-stage amplifier on the millikelvin chamber, 
such as Josephson parametric amplifiers\cite{9} during the whole experiment. 
This presents a significant advantage in reducing device costs and achieving large-scale integrations.

In conclusion, we have developed an innovative readout architecture capable of achieving a measurement fidelity of 99.8\%, 
and an estimated pure measurement fidelity above 99.9\% for superconducting qubits, 
even in the absence of any first-stage microwave amplification. 
Moving forward, the readout fidelity can be further enhanced by fine-tuning the device parameters, 
such as the Josephson energy of CJJ and the coupling quality factor of the resonator.

\emph{Note}: While preparing this manuscript, 
we became aware of a different approach using on-chip notch filters to achieve high readout fidelity (ref.\cite{28}).

$^{*}$C. Wang, and F. -M. Liu contributed equally to this work.

$^{\dagger}$Corresponding author.

~\\

\noindent \textbf{Competing interests:} 
Some of the technologies in this article have been granted a patent (ZL 202110847640.6, application date: 23-07-2021), part of the authors are inventors.

~\\

\bibliographystyle{apsrev4-1}
\bibliography{ref_num.bib}

\clearpage

\end{document}